\documentclass{iaus}
\usepackage{graphicx}

\title[Disc-Jet Connection] 
{The Disc-Jet Connection}

\author[Pudritz \& Banerjee]   
{Ralph E. Pudritz 
 \and Robi Banerjee}

\affiliation{Department of Physics \& Astronomy, McMaster University,
Hamilton, ON L8S 4M1, Canada \break emails: pudritz@physics.mcmaster.ca;
banerjee@physics.mcmaster.ca\\[\affilskip]}

\pubyear{2005}
\volume{227}  
\pagerange{119--126}
\date{?? and in revised form ??}
\jname{Massive Star Birth: A Crossroads
of Astrophysics }
\editors{E.B. Churchwell, M. Felli \& C.M. Wadsley, eds.}
\begin{document}

\maketitle

\begin{abstract}

A large body of theoretical and computational work 
shows that jets - modelled as magnetized disk winds -
exert an external torque on their underlying disks that
can efficiently remove angular momentum and 
act as major drivers of
disk accretion.
These predictions 
have recently been confirmed in direct HST measurements of
the jet rotation and angular momentum transport 
in low mass protostellar systems.
We review the theory of disc winds
and show that their physics is  
universal and scales to jets from both low and  
high mass star forming regions.  
This  
explains the observed properties of outflows in 
massive star forming 
regions, 
before the central massive star
generates an ultracompact HII region. 
We also discuss 
the recent numerical studies on the formation of
massive accretion disks and outflows 
through gravitational collapse, including 
our own work on 3D Adaptive Mesh simulations (using
the FLASH code) of the hydromagnetic 
collapse of
an initial rotating,
and cooling Bonner-Ebert sphere. 
Magnetized collapse gives rise to outflows.
Our own simulations show that
both a jet-like disk wind on sub AU scales,  and a larger scale 
molecular outflow occur (Banerjee \& Pudritz 2005). 

\end{abstract}

\firstsection 
\section{Introduction}

The formation of massive stars is one of the most important
areas of star formation research because of its broad 
implications for many different aspects of astrophysics - 
from the role that the first stars played in ending the cosmic dark 
age to the control that massive stars exert on   
galactic evolution.  Of the two current theoretical models for
the formation of massive stars - agglomeration through
stellar collisions (eg. Bonnell et al. 1998) or accretion through a disk 
(eg. Yorke and Sonnalter 2002), the very earliest
stages in the latter picture is a scaled
up version of star formation as it is observed for lower mass
stars.  

In the accretion picture,   
most of the 
collapsing material in a massive and probably turbulent
core forms a disk through which high disk accretion 
rates quickly assemble
a massive star.  In this early phase, and before
an intense stellar
radiation field turns on,    
it is natural to expect that jets
and outflows will be driven from massive,
magnetized disks (Blandford and Payne 1982, Pudritz and Norman 1983).
These flows would be expected to be governed by the
same physical principles as their low mass, TTS counterparts,
and should pre-date the appearance of the compact HII region.
We review the observational, theoretical, and computational
basis for understanding the bipolar outflows in massive star 
forming regions in the following sections.

\section{Observational clues - low vs high mass outflows and jets}

Measurements of the thrust associated with 
molecular outflows 
provide an important body of evidence that 
low and high mass outflows have a common origin. 
One finds that CO outflows span nearly 5 decades in thrust F
(units $10^{-4} M_{\odot} km/s \ yr^{-1}$) and have orders of magnitude
more thrust than can be accounted for by thermal or radiative
drives.  While there is a broad scatter, there is a clear 
relation between thrust and bolometric luminosity of the central
source (Cabrit \& Bertout, 1992);  
\begin{equation}  
F_{outflow}/F_{rad} = 250 (L_{bol}/10^3 L_{\odot})^{-0.3}.  
\end{equation}
This has been confirmed by the analysis of data
from over 390 outflows, ranging up to $10^6L_{\odot}$ in
central luminosity (Wu et al 2004).  

CO outflows in regions of high mass star formation were
first studied by Shepherd \& 
Churchwell (1996).   
High resolution observations have since determined 
the collimation factors of these outflows which can be as 
high as 10, suggesting that these are true bipolar outflows
and resemble those seen in low mass systems 
(eg. Beuther et al 2002). 
There are also a few clearcut cases of 
hot cores that have no centimeter-wave radio transmission
yet which have massive outflows.  Clearly, bipolar outflow 
precedes the appearance of an ultracompact HII
region (Gibb et al 2004).  

CO outflows in low mass systems are generally 
thought to trace the interaction
between an underlying jet and the surrounding molecular gas
(eg. reviews Cabrit et al 1997, Richer et al 2000).  
The coupling between the jet and the 
outflow has been modelled in two different ways:
either as a jet-driven bow-shock (eg. Raga \& Cabrit 1993,
Masson \& Chernin 1003) or as a wide-angle, X-wind driven 
shell (see review Shu et al. 2000).  The observations for
low mass systems suggest
that both types of interaction may occur 
depending upon the source (Lee et al. 2000).  

Discs in massive star forming regions
are difficult to detect, but
are present in at least some massive systems
as early
high spatial resolution observations showed 
(eg. Cesaroni et al. 1997, Zhang et al. 1998).  One of the
most interesting recent cases is the disk in the Omega
Nebula (M17) wherein a flared disk of up to
$M_{disc} \simeq 100 M_{\odot}$ may be present
(Chini et al. 2004).  

What drives these outflows?  While
jets are  still very difficult to detect in high mass outflows,
direct observations of  
jet dynamics and jet-disc coupling
have occurred for low mass systems.  There are now two strong
lines of observational evidence that 
support the theory that jets are collimated disc winds:
(i) jet rotation and (ii) link between disk accretion and 
outflow.  

High spatial and spectral resolution HST observations of 
emission line profiles of jets such as DG Tau clearly show
line asymmetries on either side of the jet axis
(eg. Bacciotti et al. 2002, Coffey et al. 2004).  These
can be interpreted as arising from the rotation of jets
at 0.5" (or 110 AU) from the axis, at a speed of 6-15
km per sec.  This rotation speed cannot arise from the innermost
regions of the disk as predicted by X-wind models - 
the rotation speed would be much higher than observed in that case. 
Detailed fits (eg. Anderson
et al. 2003) argue instead that the rotating material arises from 
an extended region of the disk ranging from 0.3 - 4.0 
AU from the central star. 

The link between disc accretion and mass outflow is well
documented in many observational studies (Hartmann \& Calvet 1995;
review, Calvet 2003).
For TTS,
whether the system is accreting at a low rate of
$\dot M_a \simeq 10^{-8} M_{\odot} yr^{-1}$, or in 
the FU Ori outburst state wherein $\dot M_a \simeq
10^{-4} M_{\odot} yr^{-1}$, the ratio of wind mass
loss rate to disc accretion remains the same at about
$\dot M_w / \dot M_a \simeq 0.1$.  
This disc-jet connection was predicted in early models of disc winds, and
shown to be a direct consequence of the angular momentum
equation and the efficiency of a magnetized wind to 
extract angular momentum from the disc.
We turn to these ideas next. 

\section{Theory of disc winds - universal scaling}

We consider the simplest possible description of a magnetized, 
rotating, object threaded by a large-scale field
(see reviews; 
K\"onigl \& Pudritz 2000 (KP), Pudritz 2003).  
The equations of stationary, ideal MHD are the conservation
of mass (continuity equation);
the equation of motion with conducting gas of
density $\rho$ undergoing
gas pressure ($p$), gravitational (from the central
object whose gravitational potential is $\phi$),
as well as the Lorentz forces 
(from the field ${\bf B}$); the induction equation
for the evolution of the magnetic field in the moving
fluid; as well as the conservation of magnetic flux.
These equations were written down by Chandrasekhar, Mestel,
and many others. 

One of the most interesting insights into the physics
of magnetized outflows comes by 
from the conservation of mass and magnetic flux along a field line.
Combining these,   
one can define a function
$ k$ that is a constant along a magnetic field line and
that we call the ``mass load'' of the wind;
\begin{equation}
\rho {\bf v_p} = k {\bf B_p}
\end{equation}
\noindent
This function represents the mass load per unit time, 
per unit magnetic flux
of the wind. Its value is preserved along each
field line emanating from the rotor 
(a disk
in this case) and its value is determined
on the disk surface.  It can be more revealingly written as 
\begin{equation}
k = { \rho v_p \over B_p} = {d \dot M_w \over d \Phi}
\end{equation}
The mass load profile (as a function
of the launching point of the wind) 
is determined by the physics of the
underlying disc and plays an important boundary
condition for all
aspects of jet physics.

\begin{figure}
 \includegraphics[height=6in, width=5in]{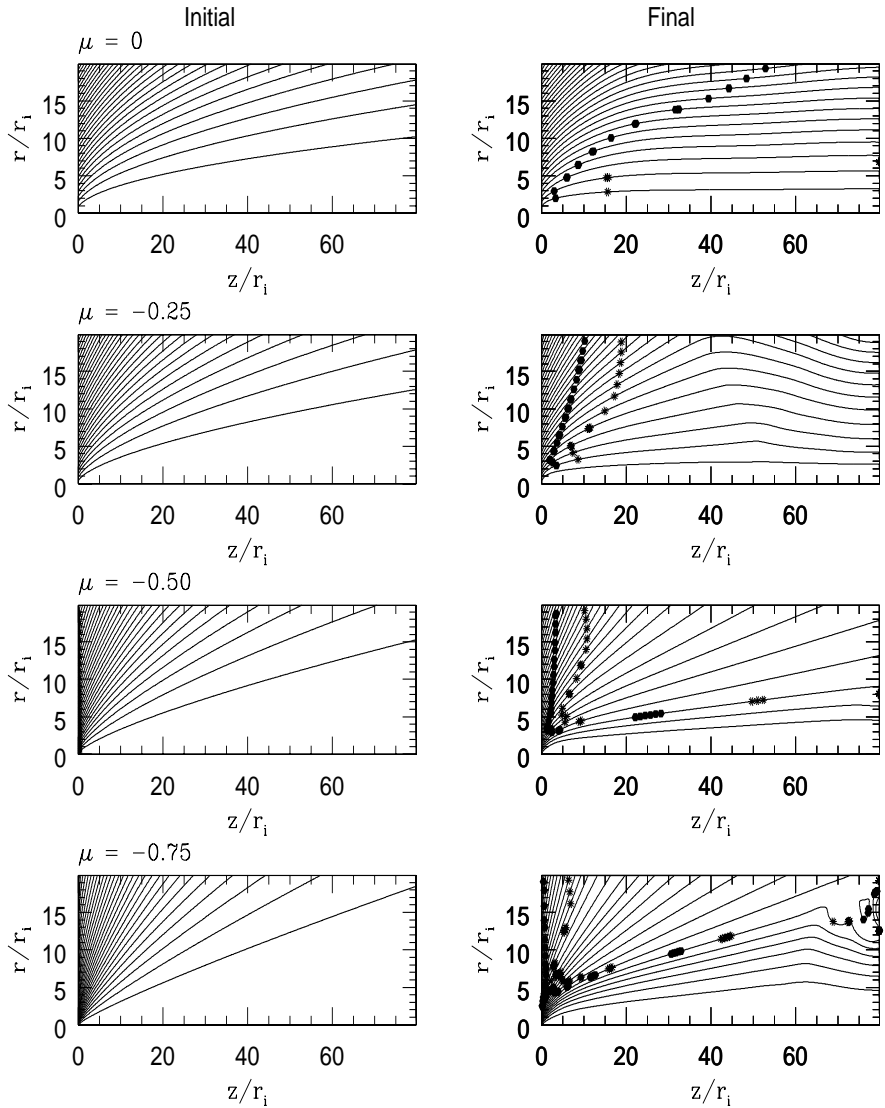}
  \caption{ Initial (left panels) and final (right 
panels) magnetic configurations for 
   disk winds for 4 models of the disc magnetic field.
   From top to bottom these are the potential solution, BP,
    PP, and a steeper $\mu = -0.75$ configuration (see equation
     3.12). Outflows are less well collimated as one goes
     from top to bottom. (Adapted from PRO05).}\label{fig:wave}
\end{figure}

The toroidal field in rotating flows follows from the induction
equation;  
\begin{equation}
B_{\phi} = {\rho  \over k} ( v_{\phi} - \Omega_o r).  
\end{equation}
where 
$ \Omega_o$ is the angular
velocity of the disk at mid-plane.
This result 
shows that the strength of the toroidal field in the jet
depends on the mass loading as well as the jet density.
At high densities, one expects a stronger toroidal field.
However, higher mass loads imply lower toroidal field 
strengths.  Jet collimation depends on hoop stress
through the toroidal field and because of this,
the mass load has a very important effect on 
jet collimation (Ouyed \&
Pudritz 1999).

The angular momentum per unit mass can be deduced from 
the condition of angular momentum conservation along each
field line; 
\begin{equation}
l = r v_{\phi} - {r B_{\phi} \over 4 \pi k} = const. 
\end{equation}
The form for $l$ reveals that the total angular momentum
is carried by both the rotating gas (first term) as well
by the twisted field (second term), the relative proportion
being determined by the mass load.

\begin{figure}
  \includegraphics[height=6in, width=5in]{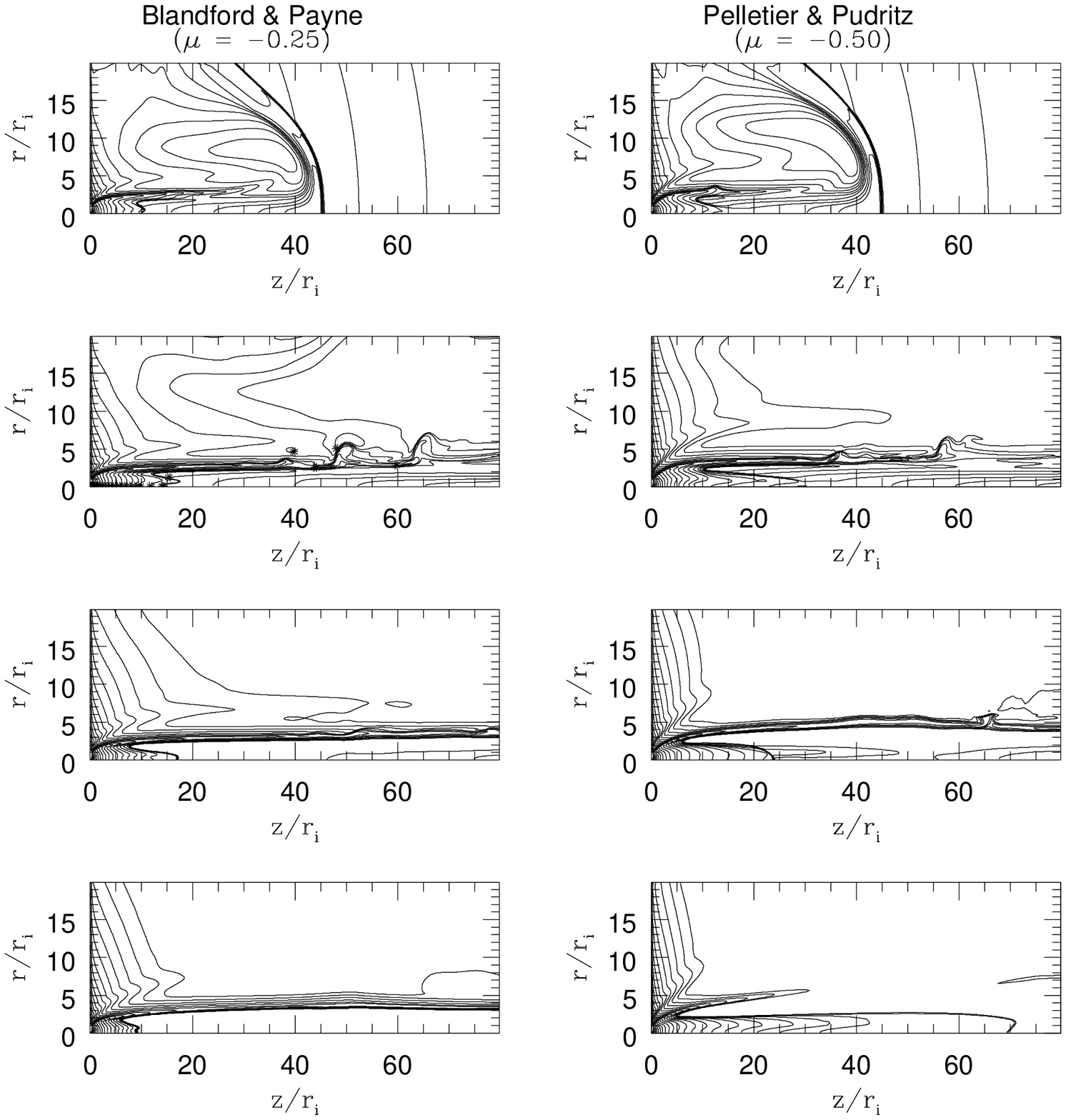}
    \caption{Four snapshots of the density structure of 
        two jets shown in 
      Figure 1:  Blandford \& Payne
      (left panels) vs. Pelletier \& Pudritz (right panels).
       The jet density peaks for material moving close to the axis.}
       \label{fig:wave}
\end{figure}

The amount of angular momentum per unit mass that is 
transported along each field line is fixed by the
position of the Alfv\'en point in the flow - where the
poloidal flow speed reaches the Alfv\'en speed for the 
first time 
($m_A=1$). It is easy to show that
the value of the specific angular momentum is;
\begin{equation}
l(a) = \Omega_o r_A^2 = (r_A/r_o)^2 l_o
\end{equation}
For a field line starting at a point $r_o$ on the
rotor (disk in our case), the Alfv\'en radius is
$r_A(r_o)$ and constitutes a lever arm for the flow.
Clearly, angular momentum is being extracted from the
rotor such that the angular momentum per unit mass
in the outflow is a factor of $(r_A/r_o)^2$ greater than
it is for gas particles in the disk ($l_o$).

\begin{figure}
\includegraphics[height=4in,width=5in]{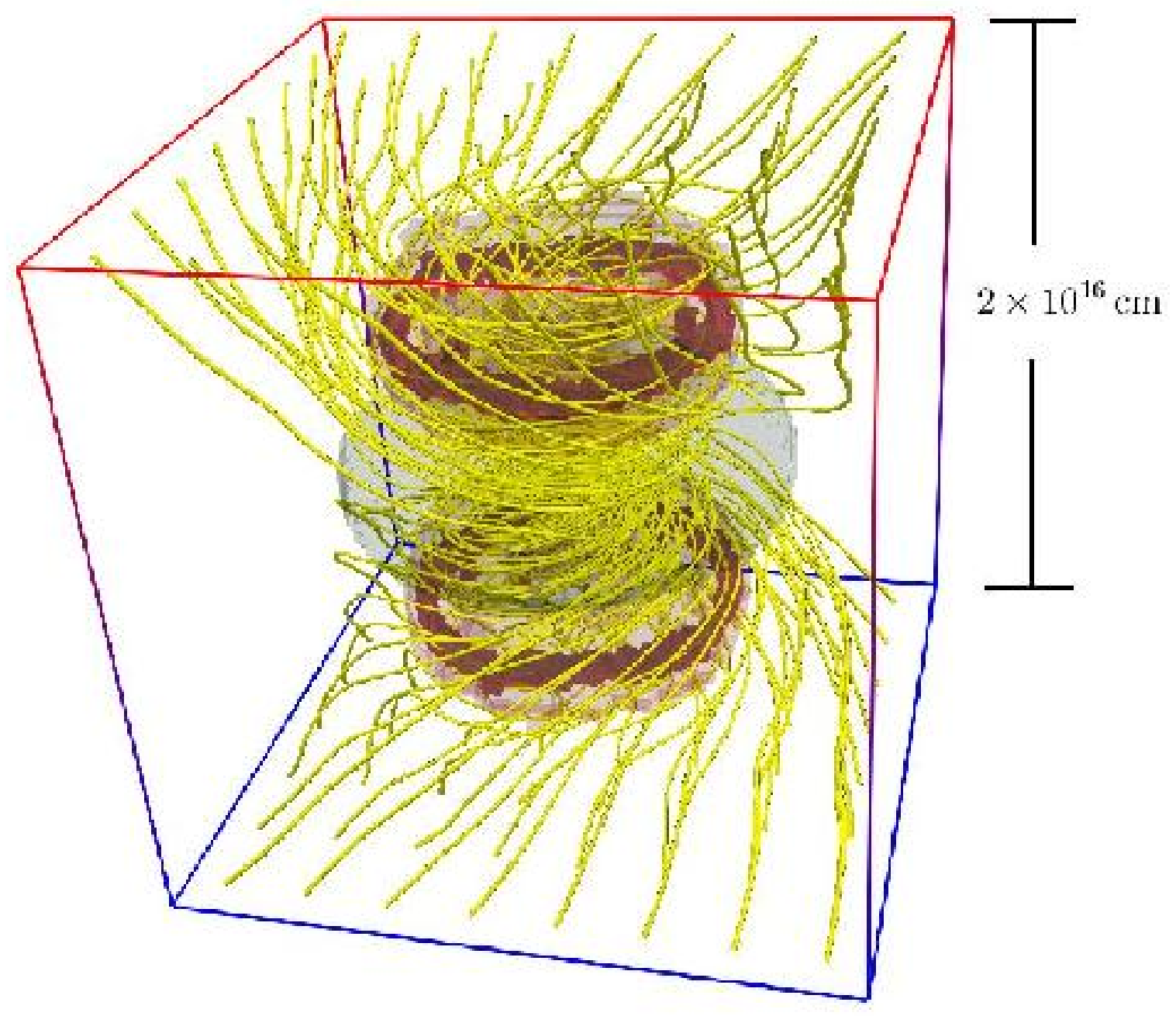}
  \caption{ Magnetic field line structure of large-scale
outflow from disc.  Isosurfaces refer to velocities 0.2
and 0.4 km s$^{-1}$. (Adapted from BP05).} 
\end{figure}

The terminal speed $v_p = v_{\infty}$ is such that the
gravitational potential and rotational energy of the flow
are negligible.  For cold flows, the pressure may also
be ignored.  It follows that
\begin{equation}
v_{\infty} \simeq 2^{1/2} \Omega_o r_A = (r_A/r_o) v_{esc,o}. 
\end{equation}
We see that the terminal speed depends on the depth
of the local gravitational well at the footpoint of the 
flow, and is therefore scalable.   

The angular momentum equation for the accretion disk undergoing 
an external magnetic torque may be written: 

\begin{equation}
\dot M_a { d (r_o v_o) \over dr_o} = - r_o^2 B_{\phi} B_z \vert_{r_o, H}. 
\end{equation}
\noindent
This result shows that angular momentum is extracted
from disks threaded by magnetic fields.  

This equation can be  
be cast into its most fundamental form;
\begin{equation}
\dot M_a {d( \Omega_o r_o^2) \over dr_o} = {d \dot M_w \over d r_o}
                   \Omega_o r_A^2
                   . (1 - (r_o/r_A)^2)
\end{equation}
which reveals that there is a crucial link between the
mass outflow in the wind, and the mass accretion rate
through the disk.  In order of magnitude, this is 
\begin{equation}
\dot M_a = (r_A/ r_o)^2 \dot M_w
\end{equation}

The value of the Alfv\'en lever arm $r_A / r_o \simeq 3$
for most theoretical and numerical models that
we are aware of, so that one finds $\dot M_w / \dot M_a \simeq 0.1$.
Observations of DG Tau suggest that $r_A/r_o \simeq 1.8 - 2.6$
(Anderson et al 2003).

\begin{figure}
\includegraphics[height=4in,width=5in]{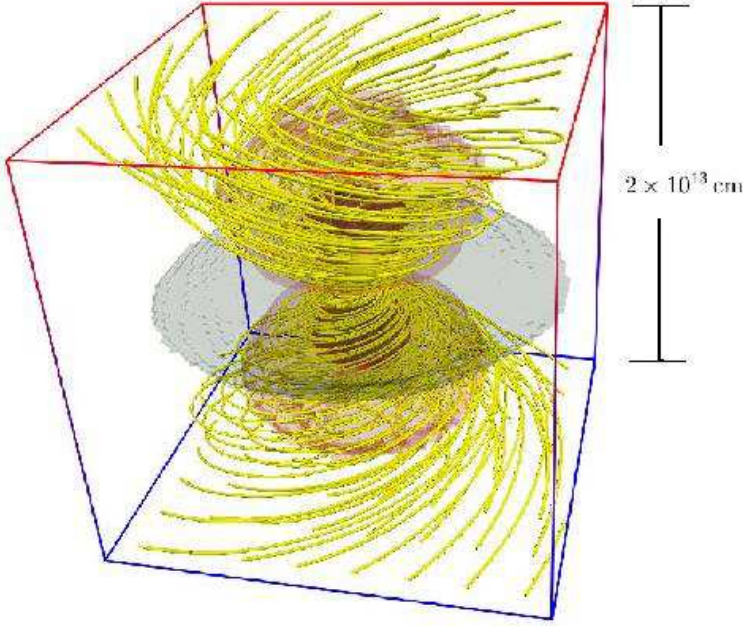}
  \caption{
Magnetic field line structure of small-scale
jet from disc.  Isosurfaces refer to velocities 0.6 
and 2 km s$^{-1}$. (Adapted from BP05).} 
\end{figure}


\begin{figure}
\includegraphics[height=4in,width=4in]{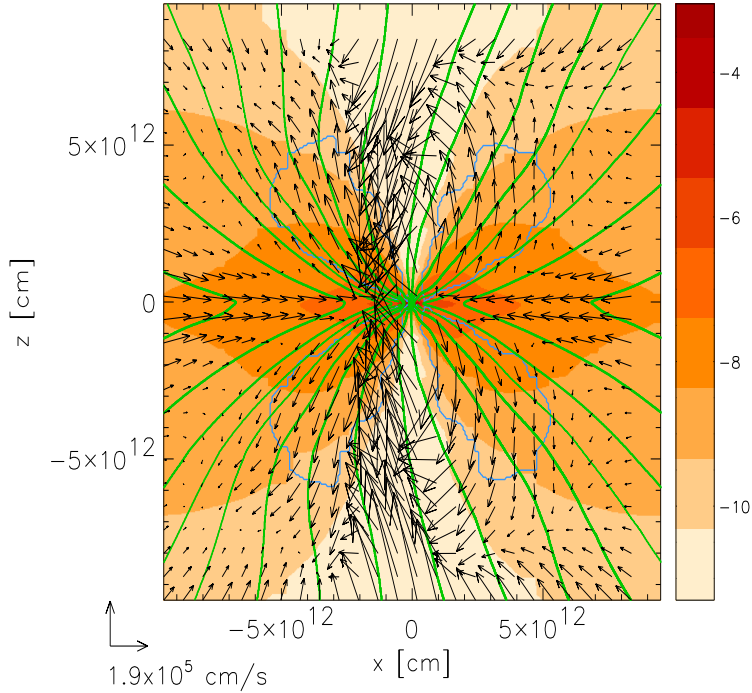}
   \caption{Cross section through the disc showing the launch
of a centrifugally driven wind that impacts envelop material
that is still infalling at this early time.
    (Adapted from BP05).}\label{fig:contour}
\end{figure}

\begin{figure}
\includegraphics[height=3in,width=4in]{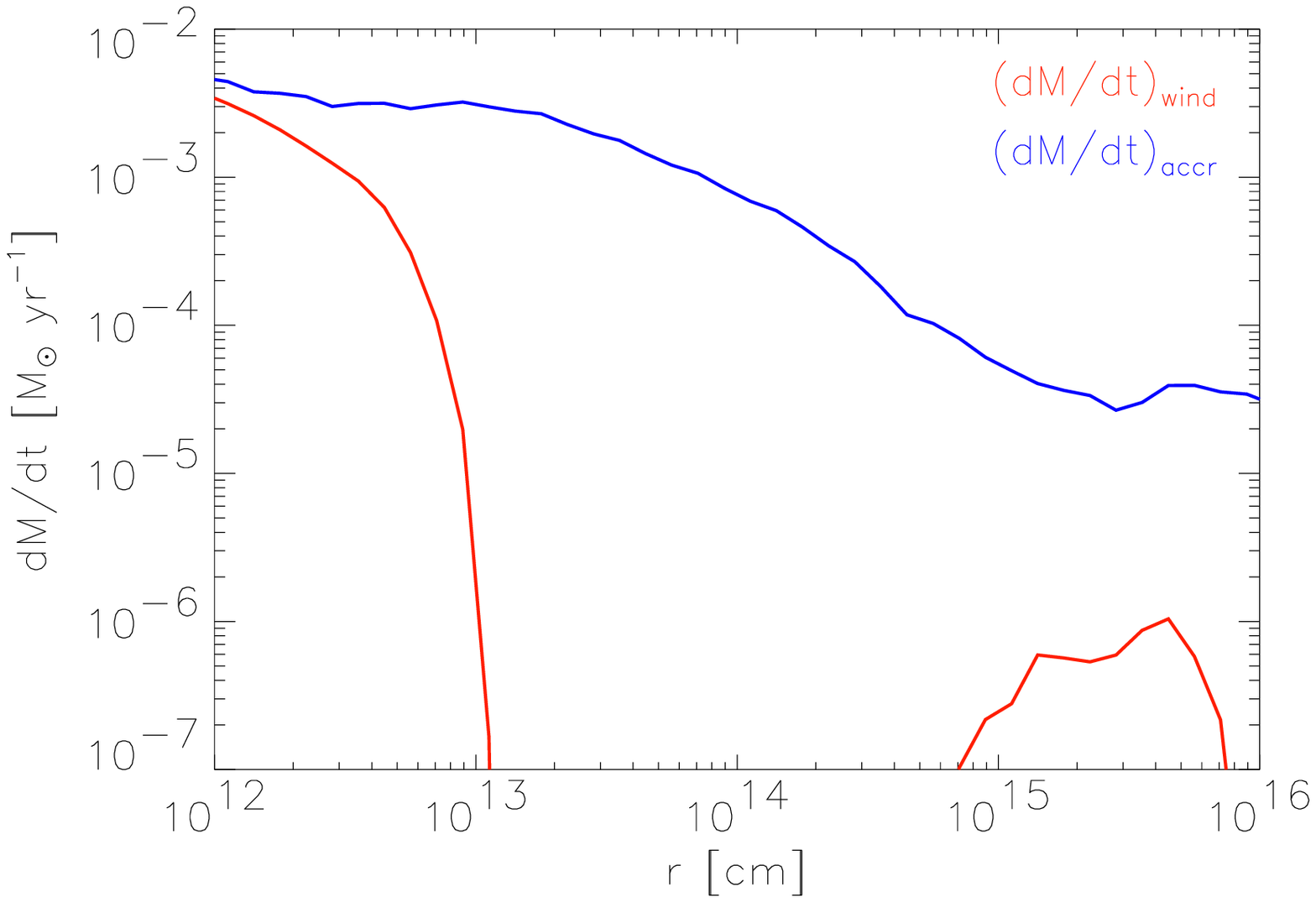}
  \caption{Accretion rate vs. mass loss rate in the outflow.
    (Adapted from BP05).} 
   \label{fig:contour}
\end{figure}

We may now connect this theoretical analysis with
the observations of momentum and energy transport in the molecular
outflows reviewed in \S 2.
The total mechanical energy that is carried
by the jet may be written as; 
\begin{equation}
L_w= {1 \over 2} \int_{r_i}^{r_j} d\dot M_w v_{\infty}^2 \simeq {1 \over 2}
                       {G M_* \dot M_a \over r_i}
                   [1 - (r_i/r_j)^2] \simeq {1 \over 2} L_{acc} .
\end{equation}
This results states that the wind luminosity taps the gravitational
energy release through accretion in the gravitational potential of
the central object.  

We may also calculate the total momentum flux (thrust) carried by
a jet launched from an accretion disk.  For simplicity, assume
that the Alv\'en radius of flow along a field line whose footpoint
is located at $r_o$ obeys a power law scaling
$r_A(r_o)/r_o \propto r_o^{-\alpha}$.  The thrust is then
\begin{equation}
F_w  = \int_{r_i}^{r_j} d \dot M_w
               v_{\infty} = [f(\alpha)/2] 
               {r_i \over r_{A,i}}
                 \dot M_a v_{esc,i} 
\end{equation}
where  $ f(\alpha) \equiv  [(1/2) + \alpha]^{-1}$ 
the index i denotes the inner disk radius, and $v_{esc,i}$
is the escape speed from $r_i$.
Again, one sees that the thrust is almost as high as it 
could theoretically be, and scales again with the depth
of the gravitational well at the inner edge of the 
accretion disk.  

A hydromagnetic wind
collimates   
because of the increasing
toroidal magnetic field in the flow as it moves
through its various critical points and away from the disk.
Collimation is achieved
by the tension force associated with the toroidal field
which leads to a radially inwards directed component
of the Lorentz force (or "z-pinch");
$ F_{Lorentz, z} \simeq J_z B_{\phi}$
where
${\bf
J} = (c/4 \pi) {\nabla \times B}$ is the current density.
Heyvaerts and Norman (1989) 
show that two types of solution for jet collimation 
are present depending upon
the asymptotic behaviour of the total current intensity $I =
2 \pi \int_0^r J_z(r',z')dr' = (c/2)
r B_{\phi}$.
In the limit that $I \rightarrow 0$ as
$r \rightarrow \infty $, then the field lines are paraboloids
which fill space.  On the other hand, if the current
is finite in this limit, then the flow is collimated to cylinders.
Given that the toroidal field in the jet depends on the
mass load, we see that the character of the flow therefore depends upon
the boundary conditions on the disk.  The similarity of collimation
factors of low and high mass outflows may suggest that 
the physics of mass loading may be similar in low and high
mass accretion discs as well.

These results 
have be rigourously tested by simulations of jets from thin disks
by many authors (see KP). In Figures 1 and 2, we show jet simulations
from the surfaces of discs that are characterized by different
mass load functions $k(r_o)$ across the disc surface
(see Pudritz 2003, PRO 2005).
We specify the surface $(r_o, 0)$
to have a power-law form
$B_z(r_o, 0) = br_o^{\mu - 1}$ and to exert no force on 
an initial corona, modelled as a gas with $\gamma = 5/3$.
The solution for $\mu = 0$ is analytic
and is
the so-called
"potential" configuration.  Other values
correspond to the  
BP model ($\mu = -1/4$),
Pelletier \& Pudritz (1992, PP92 ($\mu = -1/2$), 
and a yet more steeply declining
magnetic field such
as $\mu = -3/4$.  These intial configurations are plotted out
in the left panels of Figure 1.

The injection speed $v_{\rm inj}$ of
the material from the disk into the base of the
corona - the injection speed - which is scaled  
with the local Keplerian velocity at each radius
of the disk;  $v_{\rm inj} = 10^{-3} v_K$ at any point
on the disk. Taken altogether, 
the mass loading at the footpoint of each field line on the disk
takes the form;

\begin{equation}
k(r_o) = \rho_o v_{p,o}/B_{p,o} \propto r_o^{- 3/2} r_o^{-1/2}/r_o^{-1 + \
mu}
 \propto r_o^{-1-\mu}
\end{equation}

The right panels of Figure 1 show the final magnetic configurations. 
While the potential and BP cases achieve cylindrical collimation
on the grid, the PP one is less well collimated while the last case presents
a wide-angle wind.  In Figure 2, we compare the density evolution
at different times of the BP outflow (left) with the PP case.  This
Figure shows that the densest material in the jet does moves fairly
parallel to the outflow axis.  Figures 1 and 2 together show that
disk winds can have both a wide-angle, as well as a cylindrically
collimated aspect, depending upon the mass loading profile provided
by the 
accretion disc, and is not a unique property of X-wind models alone.    
We see that one can have wide-angle, or cylindrically collimated
flows that could explain the variety of observations noted in \S 2
(see PRO05 for details).

\section{Outflow and jet formation during gravitational collapse} 

Outflows and jets are the earliest manifestations of star formation
in both low and high mass systems.  Therefore, they must arise 
in the early stages of gravitational collapse soon after the 
underlying accretion discs begin to form.  What ingredients are
necessary for outflow to occur?

Decades worth of purely hydrodynamical
simulations of gravitational collapse in low
mass systems have shown 
that outflows do not occur in 
such purely hydrodynamic collapse. 
The initial conditions for star formation have 
become much better defined with time.  As an example,
there are now excellent observations of low mass hydrodynamic
cores that are very well fit by Bonner-Ebert density profiles
(eg. Barnard 68 with 2.1 $M_{\odot}$, Alves et al. 2001; B335
with 14 $M_{\odot}$, Harvey et al. 2001). 

We have recently 
carried out 3D simulations of the collapse of cooling, rotating
low (Barnard 68) and high mass (160 $M_{\odot}$) B-E spheres
(Banerjee et al 2004) using an Adaptive Mesh Refinement
(FLASH) code.  This allows us to resolve the local Jeans length
in the system by as many as 12 pixels, with final results that
resolves over 7 decades in length scale and can resolve down to 
0.3 AU.  Free-fall, isothermal collapse continues until the
molecular cooling time is slower than the free-fall time - which
occurs at a characteristic density of $10^{7.5}$ cm$^{-3}$.
The gas undergoes a shock, and afterwards undergoes a second collapse
onto the underlying disk.  
We observe that accretion through the disk is driven 
by a bar mode that manifests itself if the initial spin of the
core obeys $\Omega t_{ff} \simeq 0.1$.  For faster spin rates,
the discs fragment into binary systems. Accretion rates through
the discs in our high mass simulation reach peak values of
$\simeq 10^{-3} M_{\odot} yr^{-1}$.

Theoretical work over the last 2 decades has emphasized that
magnetic fields are critical for the formation of outflows and
jets.  The first simulations of the collapse of magnetized cores
and the onset of jets were produced by Tomisaka (1998).  
Using a nested grid scheme, his calculations followed
the collapse of a section of a magnetized cylinder and
showed the onset of a large scale
outflow and a jet on smaller scales (10 stellar radii typically).

Recently, we extended our purely hydrodynamic B-E collapse
calculations to include a uniform, threading magnetic field.
The change in dynamics is remarkable because
we observe the onset of a large
scale outflow on 100 AU scales as well an inner jet 
on sub AU scales (Banerjee \& 
Pudritz 2005 (BP05)).   In Figure 3 we
show a 3D visualization of a large scale outflow 
that
begins at
about 70,000 yrs after the initiation of the collapse
of the Barnard 68 low mass systems.  
It erupts from the surface of the disc at about
100 AU, and extends up to the outer accretion shock.  By
the end of the simulation it is still propagating outwards
away from the disc driving a shock front ahead of it.  
It has the character of a slow,
outward propagating magnetic tower that is driven by the 
gradient of the twisted toroidal field that is 
attached to the disc (Lynden-Bell 2003).  
This outflow 
is clearly dominated by a strongly
wrapped toroidal field that pushes a ring of gas outwards and
away from the disk.

The jet that arises on smaller scales is a true disc
wind and it achieves super-Alfv\'enic speeds.  It is shown in 
a 3D visualization in Figure 4, where we see the wrapped
toroidal field lines that provide the collimating hoop stress 
on the jet.  The jet has not yet burst out of the infall region.  
In Figure 5, we present an x-z plane section 
of the disc that is 
perpendicular to the disc midplane. The 
magnetic field lines in the disc are strongly 
bent away from the vertical axis by the accretion flow in the
disc, which nicely sets up the initial condition for the 
centrifugal launch of the jet.  

Finally, we show the mass tranport rate in both the outflow
(bottom, outer curve), and the
inner jet (bottom, inner curve), as compared to the 
accretion rate. 
(top curve).  These rates are computed as the average flow rate
through a sphere of radius r.  First, the accretion rate through
these magnetized discs have increased compared
to the pure hydrodynamic case.  The accretion
driven by the wind is comparable
to the flow driven by the spiral waves in the disc at the point
that our simulation ends. 
Second, the ratio of 
outflow to accretion rate through the inner part of the disc
where the jet is active is $\dot M_w/ \dot M_a \simeq 1/3$!  

We conclude that jets and outflows are an intrinsic part of the 
collapse of magnetized cores, and that their properties are 
independent of the mass of the system.  Both low and high mass
outflows should share the same physics.  Things change when 
the massive central star turns on its radiation field, but by
that time, the outflow has already carved a channel through the 
core that may play a crucial part in relieving the radiation 
pressure in the interior region.  This may promote the
further growth, through accretion, of the central
massive star (eg. Krumholz et al. 2005).  

\begin{acknowledgments}
We thank Rachid Ouyed and Sean Matt for many interesting
discussions on this topic. RB is a SHARCnet Fellow at McMaster
University, and REP's research is supported by 
NSERC of Canada. 
\end{acknowledgments}

\end{document}